\documentclass[amsmath,amssymb,prl,aps,showpacs,superscriptaddress,twocolumn]{revtex4}
\usepackage{amsmath,amsfonts,amssymb,amsthm,graphics,graphicx,epsfig,bbm}
\usepackage[colorlinks=true,citecolor=blue,linkcolor=blue,urlcolor=blue]{hyperref}
\usepackage[usenames]{color}
\usepackage{graphicx}
\usepackage{subfigure}
\usepackage{amsmath}
\usepackage{epsfig}
\usepackage{dcolumn}
\usepackage{bm}
\usepackage{color}
\usepackage{verbatim}
\usepackage{epstopdf}
\usepackage{amssymb} 
\usepackage{amstext}
\usepackage{latexsym}
\usepackage{hyperref}
\usepackage{amsfonts}
\usepackage{psfrag}
\usepackage{xcolor}
\usepackage{times}

\def\beq{\begin{equation}}
\def\eeq{\end{equation}}\newcommand{\beqa}{\begin{eqnarray}}
\newcommand{\eeqa}{\end{eqnarray}}

\newcommand{\psild}{\left( \dot{\psi}_i^l \right)}

\newcommand{\psird}{\left( \dot{\psi}_i^r \right)}
\newcommand{\para}[1]{\left( #1 \right)}
\newcommand{\parb}[1]{\left[ #1 \right]}
\newcommand{\parc}[1]{\left\{ #1 \right\}}

\begin{document}
\title{Dynamical Casimir effect entangles artificial atoms}
\author{S. Felicetti}
\affiliation{Department of Physical Chemistry, University of the Basque Country UPV/EHU, Apartado 644, E-48080 Bilbao, Spain}
\author{M. Sanz}
\affiliation{Department of Physical Chemistry, University of the Basque Country UPV/EHU, Apartado 644, E-48080 Bilbao, Spain}
\author{L. Lamata}
\affiliation{Department of Physical Chemistry, University of the Basque Country UPV/EHU, Apartado 644, E-48080 Bilbao, Spain}
\author{G. Romero}
\affiliation{Department of Physical Chemistry, University of the Basque Country UPV/EHU, Apartado 644, E-48080 Bilbao, Spain}
\author{G. Johansson}
\affiliation{Department of Microtechnology and Nanoscience (MC2), Chalmers University of Technology, SE-412 96 G\"oteborg, Sweden}
\author{P. Delsing}
\affiliation{Department of Microtechnology and Nanoscience (MC2), Chalmers University of Technology, SE-412 96 G\"oteborg, Sweden}
\author{E. Solano}
\affiliation{Department of Physical Chemistry, University of the Basque Country UPV/EHU, Apartado 644, E-48080 Bilbao, Spain}
\affiliation{IKERBASQUE, Basque Foundation for Science, Alameda Urquijo 36, 48011 Bilbao, Spain}

\begin{abstract} 
We show that the physics underlying the dynamical Casimir effect may generate multipartite quantum correlations. To achieve it, we propose a circuit quantum electrodynamics (cQED) scenario involving superconducting quantum interference devices (SQUIDs), cavities, and superconducting qubits, also called artificial atoms. Our results predict the generation of highly entangled states for two and three superconducting qubits in different geometric configurations with realistic parameters. This proposal paves the way for a scalable method of multipartite entanglement generation in cavity networks through dynamical Casimir physics.
\end{abstract}
\date{\today}
\maketitle

The phenomenon of quantum fluctuations, consisting in virtual particles emerging from vacuum, is central to understanding important effects in nature\textemdash for instance, the Lamb shift of atomic spectra~\cite{Lamb1947}  and the anomalous magnetic moment of the electron~\cite{QFTBook}.  
The appearance of a vacuum-mediated force between two perfectly conducting plates, known as the Casimir effect, is caused by a reduction of the density of electromagnetic modes imposed by the boundary conditions~\cite{Casimir1948,Casimir1997,Casimir1998}. This leads to a vacuum radiation pressure between the mirrors that is lower than the pressure outside. It was also suggested~\cite{DCasimir} that a mirror undergoing relativistic motion could convert virtual into real photons. This phenomenon, denominated dynamical Casimir effect (DCE), has been observed in recent experiments with superconducting circuits~\cite{CMWilson2011,Hakonen2013}. In the same manner that the Casimir effect can be understood as a mismatch of vacuum modes in space, the kinetic counterpart can be explained as a mismatch of vacuum modes in time.

A moving mirror modifies the mode structure of the electromagnetic vacuum. If the mirror velocity, ${\it v}$, is much smaller than the speed of light, ${\it c}$, then the electromagnetic modes adiabatically adapt to the changes and no excitations occur. Otherwise, if the mirror experiences relativistic motion, changes occur nonadiabatically and the field can be excited out of the vacuum, generating real photons. Beyond its fundamental interest, it has been pointed out that the DCE provides a mechanism to generate quantum correlations~\cite{Dalvit2000,Maia2000,Narozhny2001,Narozhny2003,Dodonov2005,Andreata2005,Dodonov2012}. In this sense, we may consider the study of the DCE as a resource for quantum networks and quantum simulations in the frame of quantum technologies. In circuit quantum electrodynamics, DCE photons have been created by modifying the boundary condition for the electromagnetic field~\cite{CMWilson2011}.  In a similar experiment photons have also been created by modulating the effective speed of light~\cite{Hakonen2013}. Note that the emergence of the DCE physics in a different quantum platform allows for other geometric configurations and interaction terms, leading to a variety of different physical conditions.

\begin{figure}[b]
\centering
\includegraphics[angle=0, width=0.4\textwidth]{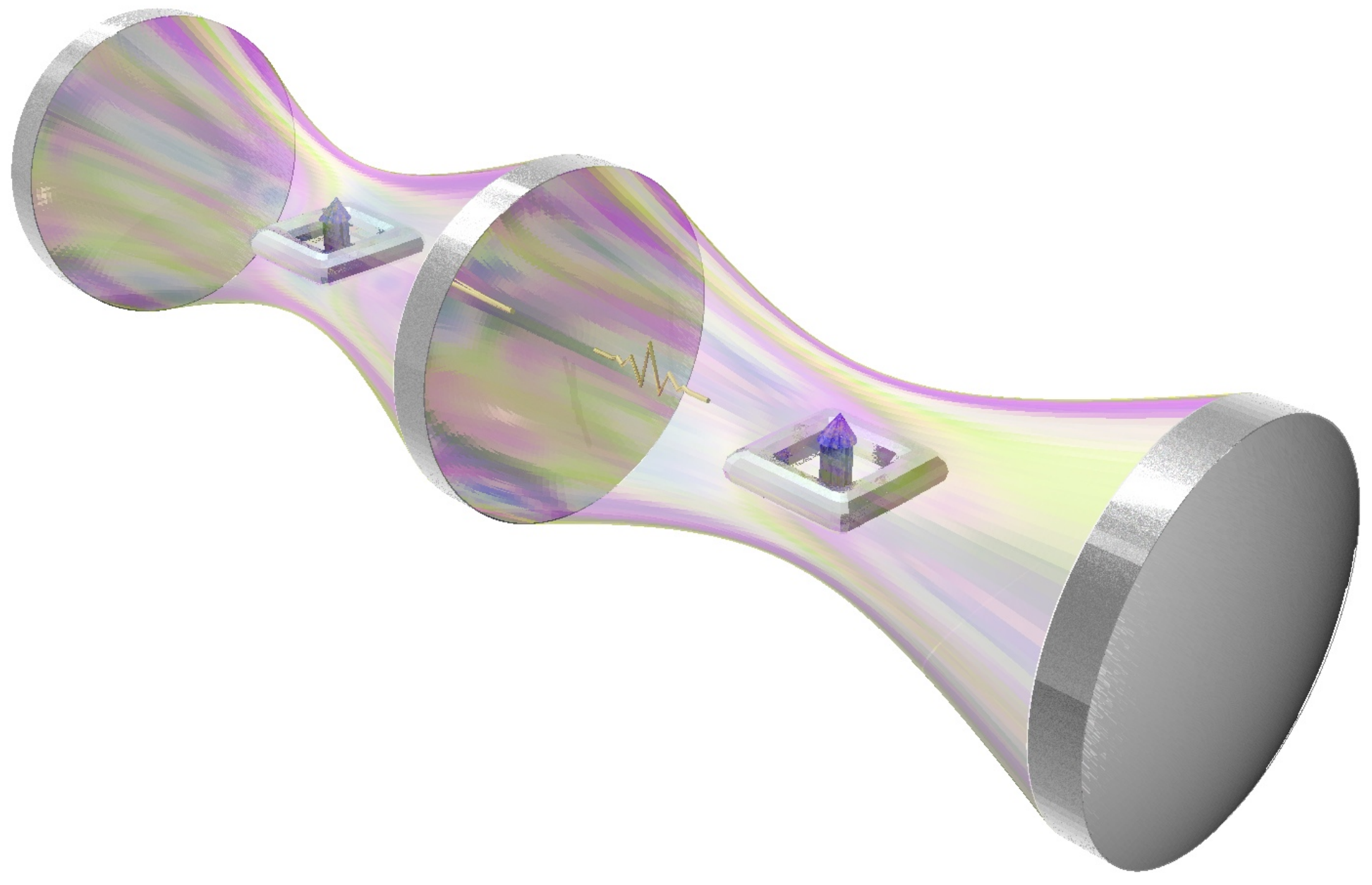}
\caption{\label{Fig1}Quantum optical implementation of the model of Eq.~\eqref{hamiltonian}:  two cavities with a common partially-reflecting mirror, each one containing a two-level artificial atom in the strong-coupling regime. If the position and/or transmission coefficient of the central mirror is time-modulated,  correlated photon pairs are generated and entanglement is transferred to qubits via the Jaynes-Cummings interaction.}
\end{figure}

In this Letter, we investigate how to generate multipartite entangled states of two-level systems, also referred  to as quantum bits (qubits), by means of varying boundary conditions in the framework of superconducting circuits. For pedagogical reasons, we illustrate our model with a hypothetical quantum-optical system, shown in Fig.~\ref{Fig1}. It is composed of two cavities that are coupled to independent single qubits. These cavities share a partially reflecting and transparent mirror, yielding the last interaction term of Hamiltonian in Eq.~(\ref{Fig1}). We assume that the cavity-qubit coupling strength is much larger than any decoherence rate in the system. In this context, we introduce the key concepts allowing the generation of highly-entangled two-qubit states, also known as  Bell states~\cite{chuang}, in circuit QED~\cite{Blais04,Chiorescu04,Wallraff04}. Later, we will consider the generation of tripartite entanglement~\cite{Acin2001} and the scalability aspects of our proposal to multipartite systems (see Supplemental Material).

The Hamiltonian describing the system of Fig.~\ref{Fig1} is composed of the sum of two Jaynes-Cummings (JC) interactions and a time-dependent coupling between the field quadratures,
\begin{eqnarray}
\label{hamiltonian}
\mathcal{H}& =&  \hbar \sum_{\ell =1}^{2}  \left[ \omega_{\ell} a_{\ell}^\dagger a_{\ell} + \frac{\omega^q_{\ell}}{2} \sigma_{\ell}^z + g_{\ell} \left( \sigma_{\ell} ^+  a_{\ell} + \sigma_{\ell}^- a_{\ell}^\dagger  \right) \right] \\
&+& \hbar \alpha (t) \left( a_1^\dagger + a_1\right) \left(a_2^\dagger + a_2 \right). \nonumber
\end{eqnarray}
Here, $a_{\ell}^\dagger$, $a_{\ell}$ are the creation and annihilation operators of the bosonic modes representing the cavity fields, while  $\sigma^z_{\ell}$, $\sigma_{\ell}^\pm$ are the Pauli operators of qubits. The characteristic frequencies of the two cavities are denoted by $\omega_{\ell}$, while the qubit energies are $\omega_{\ell}^q$. The parameters $g_{\ell}$ and $\alpha (t)$ denote the cavity-qubit and cavity-cavity interaction strength, respectively.
 
 In Eq.~\eqref{hamiltonian}, the coupling between different cavity modes, due to the overlap of their spatial distribution, is written in its full form without performing the rotating wave approximation. 
While in optical cavities this overlap can be obtained with a partially reflecting mirror~\cite{Hartmann2006}, in circuit QED it is commonly implemented using capacitors or inductances shared by two or more resonators.  
The boundary condition at the edge shared by the cavities is ruled by the central mirror position and by its reflection coefficient. Modulating these physical quantities results in a time dependence of the cavity frequencies $\omega_i$ and of the coupling parameter $\alpha$. When the effective cavity length is oscillating with small deviations from its average value, we can still consider the system as a single-mode resonator. In particular, if the cavity-cavity coupling parameter is a time-dependent function, $\alpha(t) = \alpha_0 \cos{\left( \omega_d t\right)}$ with $\omega_d = \omega_1 + \omega_2$ and $\alpha_0/\omega_i\ll 1 $, the interaction effectively turns into a two-mode squeezing term (see below),
\begin{equation}
\alpha(t) X_1 X_2 \rightarrow \frac{\alpha_0}{2} \left(a_1^\dagger a_2^\dagger + a_1a_2 \right),
\label{Twomode}
\end{equation}
which generates pairs of entangled photons shared by the cavities. By means of the Jaynes-Cummings interaction, entanglement generated between  cavities may be transferred to resonant qubits. In fact, we will prove below that, under suitably designed conditions, maximal entanglement (Bell state) between the two qubits may be attained.
\begin{figure}[t]
\centering
\includegraphics[angle=0, width=0.45\textwidth]{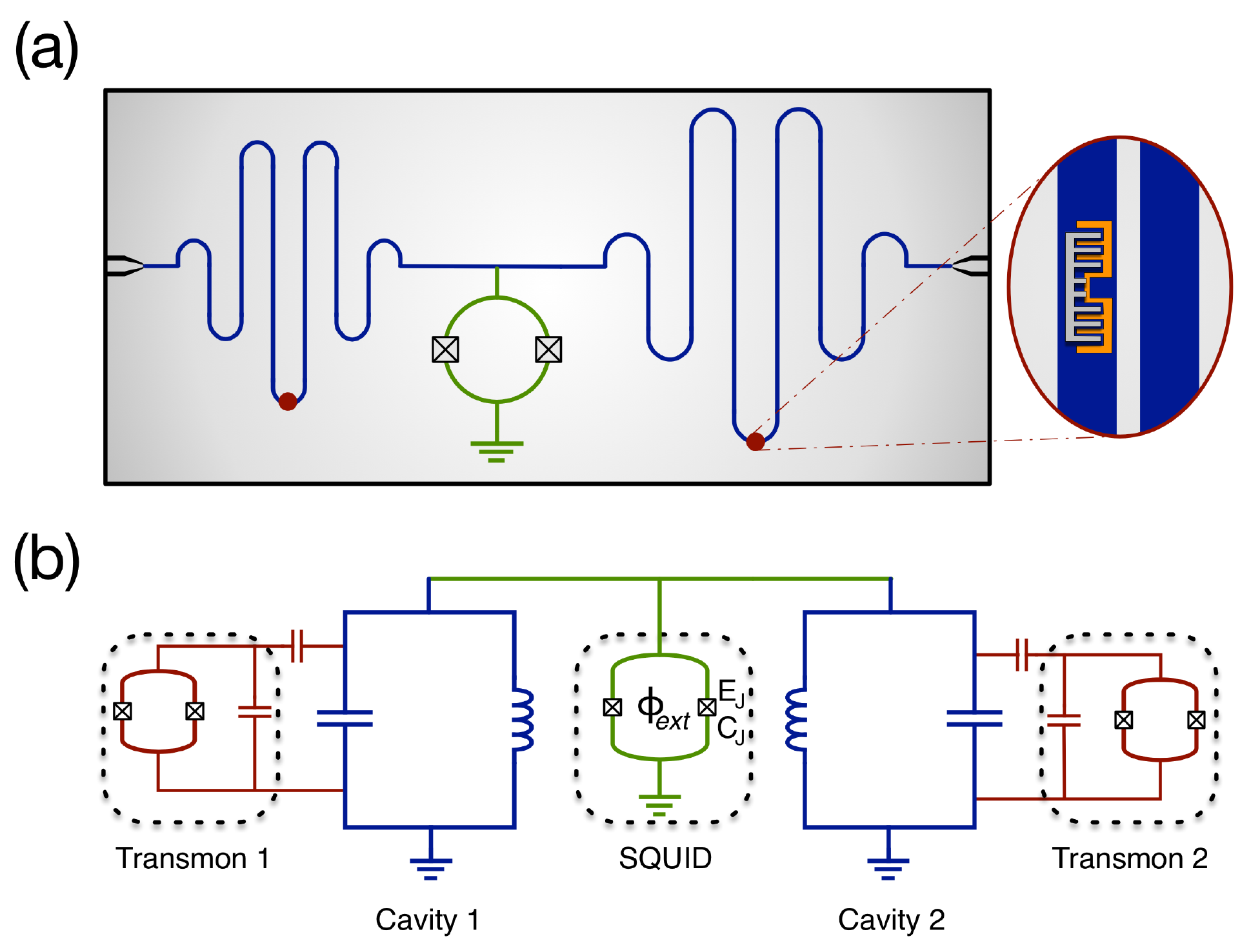}
\caption{\label{Fig2}  (a) The model of Fig.~\ref{Fig1} can be  implemented by means of two coplanar waveguides, grounded through a SQUID, containing two superconducting qubits. The blue lines represent two parallel strip lines of isolating material, where the superconducting region between them constitutes the coplanar waveguide. Each cavity interacts with a transmon qubit that is denoted by a red dot. Different resonator lengths result in distinct resonator frequencies.  (b) Circuit diagram for the previous scheme, where the cavities are effectively represented by LC resonators. We assume two identical Josephson junctions of the SQUID, while transmon qubits are constituted by two Josephson junctions shunted by a large capacitance.}
\end{figure}

Nowadays, quantum technologies~\cite{QuDevices} offer several platforms to study fundamentals and applications of  quantum theory. In particular, superconducting circuits technology~\cite{WC2008,DS2013} is a prime candidate to implement the model of Eq.~\eqref{hamiltonian}. In this framework, the cavities are constituted by coplanar waveguides, working at cryogenic temperatures, that are described by an equivalent LC circuit, as shown in Fig.~\ref{Fig2}(a,b). The characteristic frequency of such devices is in the $2-10$~GHz microwave regime. Each cavity can be coupled to a superconducting qubit built from Josephson junctions (JJs) to access charge~\cite{Vouchiat1998}, flux~\cite{Mooij1999}, or phase~\cite{Martinis1985} degrees of freedom. Specifically, we propose the use of transmon qubits which have low sensitivity to charge noise and coherence times well above ten $\mu$s~\cite{Koch2007, Paik2011, Chang2013}. 
The moving mirror~\cite{Wall2006,Bruschi2013} that couples both cavities (see Fig.~\ref{Fig1}) can be implemented by means of a superconducting quantum interference device (SQUID)~\cite{OrlandoBook}, which behaves as a tunable inductance. A SQUID is composed of a superconducting loop interrupted by two JJs (see Fig.~\ref{Fig2}(a)), threaded by an external flux $\phi_{\rm ext}$. The latter allows a fast modulation of the electrical boundary condition of cavities and their interaction. Notice that a modulation of the magnetic flux threading the SQUID induces a proportional  variation of the effective resonator lengths, while in the system of Fig.~\ref{Fig1}, moving the central mirror results in an opposite change of cavity lengths. 

\begin{figure}[t]
\centering
\includegraphics[angle=0, width=0.4\textwidth]{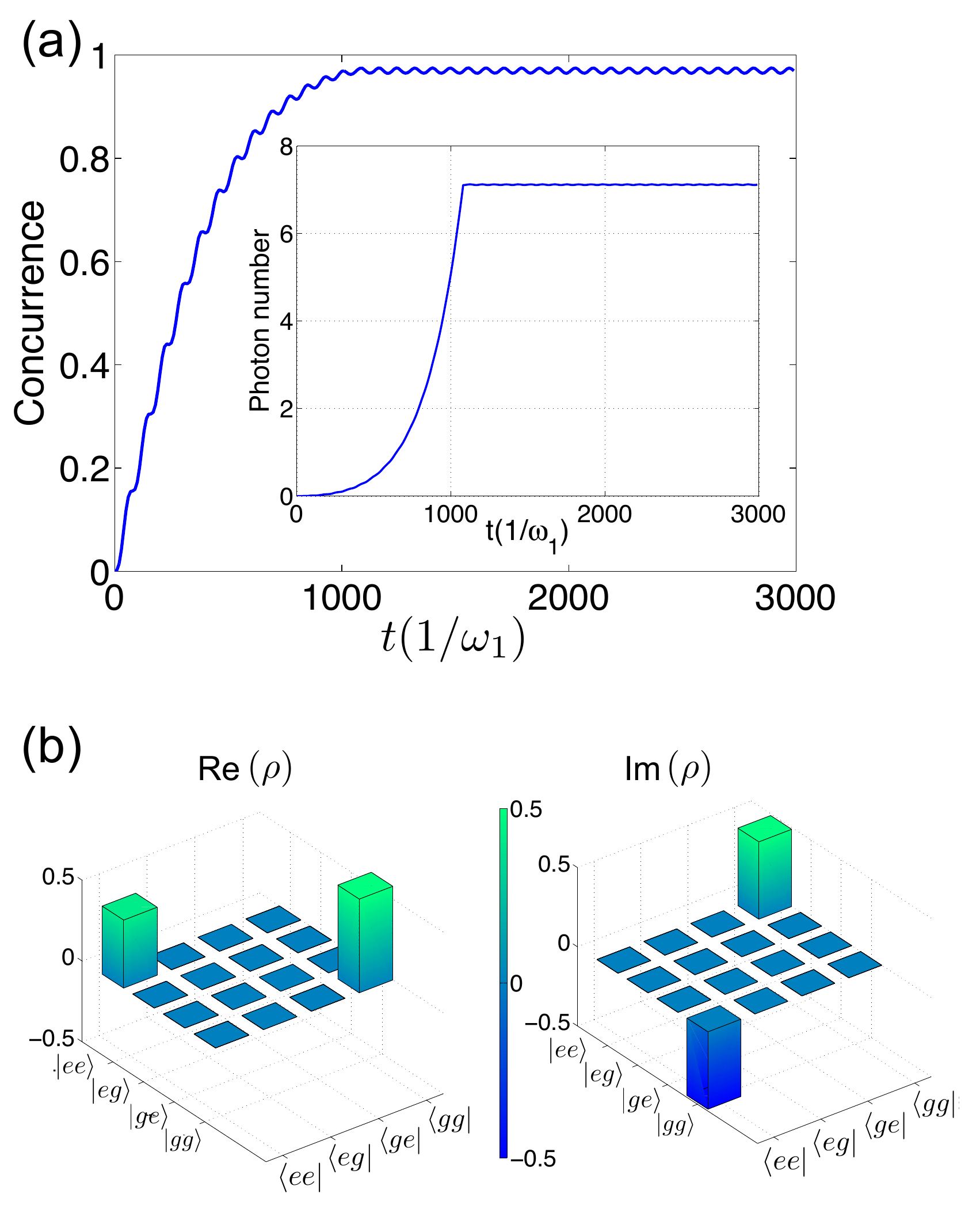}
\caption{\label{Fig3} (a) Concurrence and mean photon number as a function of time in units of the cavity frequency $\omega_1$. Here, the chosen parameters are: $\omega_1/2\pi= 4$~GHz, $\omega_2/2\pi = 5$~GHz, the impedance for both cavities is $Z_0 = 50 \Omega$, and the critical current of the SQUID junctions is $I_C = 1.1$~$\mu$A. Such parameters result in a squeezing parameter $\alpha_0=\omega_1\times 10^{-3}$. Each qubit is resonant with its corresponding cavity and they are coupled with the same interaction strength 
$g =0.04~\omega_2$.  (b) Real and imaginary parts of the density matrix $\rho$ associated with the two-qubit system.}
\end{figure}

By using off-the-shelf electronics, it is possible to produce magnetic fluxes that oscillate at the cavity characteristic frequencies. The upper limit to the speed of modulation is imposed by the SQUID plasma frequency, defined as $\omega_p=\frac{1}{\hbar}\sqrt{8E_CE_J}$, where $E_C$ is the charging energy, $E_J$ the Josephson energy, both associated with a single JJ belonging to the superconducting loop. Beyond this frequency, the internal degrees of freedom of the device are activated and a more complex behavior appears. To overcome this problem, the external flux $\phi_{\rm ext}(t)$ injected into the device, which also determines $E_J$,  will be composed of the sum of a signal oscillating at the driving frequency $\omega_d$ and a constant offset $\phi_0$, $\phi_{\rm ext}(t) = \phi_0 + \Delta \phi \cos{\left( \omega_d t \right)}$. We consider nondegenerate resonators to avoid uncorrelated photon generation at the cavity resonance frequencies, an assumption that has been confirmed by a detailed quantum mechanical analysis of the effective lumped circuit element in Fig.~\ref{Fig2}(b) (see Supplemental Material).    

If the instantaneous resonant frequency of a given resonator follows the time-dependence $\omega(t) = \omega_0 + \delta\omega \cos \left( \omega_d t \right) $,  cavity modes are well defined only under the condition $\delta \omega \ll \omega_0$. In our proposal, the frequencies of the cavity modes are obtained by solving the transcendental equation $kd\tan{(kd)} = L/L_s -C_s/C (k d)^2$ for the wave number $k$, where $d$ is the length of the resonator. We called $C_s$, $L_s$ and $C$, $L$ the effective capacitance and inductance of the SQUID and of the resonator, respectively.
Parameters used in our simulations assure that $\delta \omega/\omega_0 < 10^{-3}$.
 
 In the interaction picture, the parametric processes induced by the SQUID lead to the Hamiltonian
\begin{eqnarray}
\label{Hdrive}
\!\!\!\! \mathcal{H}^{I}_{\rm d}(t) & = & \hbar \cos({\phi_{\rm ext}/\varphi_0 })\Big[\sum^{2}_{\ell =1}\alpha_{\ell}(a_{\ell}e^{-i\omega_{\ell}t} + a^{\dag}_{\ell}e^{i\omega_{\ell}t})^2\nonumber \\
&-& \hbar \tilde{\alpha}(a_{1}e^{-i\omega_{1}t} + a^{\dag}_{1}e^{i\omega_{1}t})(a_{2}e^{-i\omega_{2}t} + a^{\dag}_{2}e^{i\omega_{2}t})\Big],
\end{eqnarray}     
where $\varphi_0=\hbar/2e$ is the reduced flux quantum, and the coefficients $\alpha_{\ell}$ and $\tilde{\alpha}$ are functions of the Josephson energy ($E_J$), the junction capacitance ($C_J$), the cavity parameters such as capacitance ($C_{\ell}$) and inductance ($L_{\ell}$). If the parameters $\alpha_{\ell}$ and $\tilde{\alpha}$ are much smaller than cavity frequencies $\omega_{\ell}$, we can perform the rotating wave approximation (RWA), and so neglect fast-oscillating terms in Eq.~\eqref{Hdrive}.
 In this case, if we consider $\phi_{\rm ext} = \phi_0 + \Delta \phi \cos{\left( \omega_d t \right)}$ with $\Delta\phi$ a small flux amplitude, the controlling the driving frequency $\omega_d$ allows to selectively activate interaction terms in the system dynamics. When the cavity is off-resonant and $\omega_d=\omega_1+\omega_2$, the interaction Hamiltonian reads as Eq.~(\ref{Twomode}).    
Interactions among different cavity modes, called mode mixing, are activated under the frequency-matching condition $\omega_d = \omega_a - \omega_b$. Cavity and driving frequencies can be chosen in order to make the relevant mode interact only with off-resonance, overdamped modes. Circuit design allows each qubit to be resonantly coupled with a single cavity mode, in which activation of higher modes due to the DCE mechanism can be neglected.

Our protocol for generating entanglement requires neither direct~\cite{Steffen2006} nor single cavity-bus mediated~\cite{DiCarlo2009} qubit-qubit interaction. Instead, it consists in cooling down the system to its ground state, turning on the external driving flux $\phi_{\rm ext}$ and switching it off at time $t_{SO}$, when the maximal qubit entanglement is reached. The {\it concurrence} $\mathcal{C}$ is an entanglement monotone of a given bipartite mixed state $\rho$, namely, the minimum average entanglement of an ensemble of pure states that represents $\rho$. For an arbitrary two-qubit state the concurrence reads~\cite{Wootters1998}
$\mathcal{C}(\rho) = \max{\left\{ 0, \lambda_1 -\lambda_2-\lambda_3-\lambda_4 \right\}}$,
where $\lambda_i$ are the eigenvalues, in decreasing order, of the Hermitian matrix $R = \sqrt{ \sqrt{\rho} \tilde{\rho} \sqrt{\rho}}$, with $\tilde{\rho}=\sigma_y\otimes\sigma_y\rho^*\sigma_y\otimes\sigma_y$.

The numerical results are shown in Fig.~\ref{Fig3}(a).  An almost maximally entangled state ($\mathcal{C}\!=\!0.97$) can be reached within $t_{SO}\!\approx\!10\!-\!500$ ns, that is, for a wide range of realistic system parameters (see Supplemental Material). 
Such protocol allows generation of the Bell state $|\psi\rangle=(|ee\rangle + i|gg\rangle)/\sqrt{2}$ with fidelity $\mathcal{F}=|\langle\psi|\rho|\psi\rangle| = 0.99$, with current superconducting circuits technology.
The density matrix of the produced Bell state is shown in Fig.~\ref{Fig3}(b). We have also proven that entanglement generation is robust against small imperfections due to limited fabrication precision and imperfect ground state preparation. Our protocol can be implemented in an on-chip architecture and it does not require any external source of squeezed signals~\cite{Kraus2014}.

\begin{figure}[]
\centering
\includegraphics[angle=0, width=0.3\textwidth]{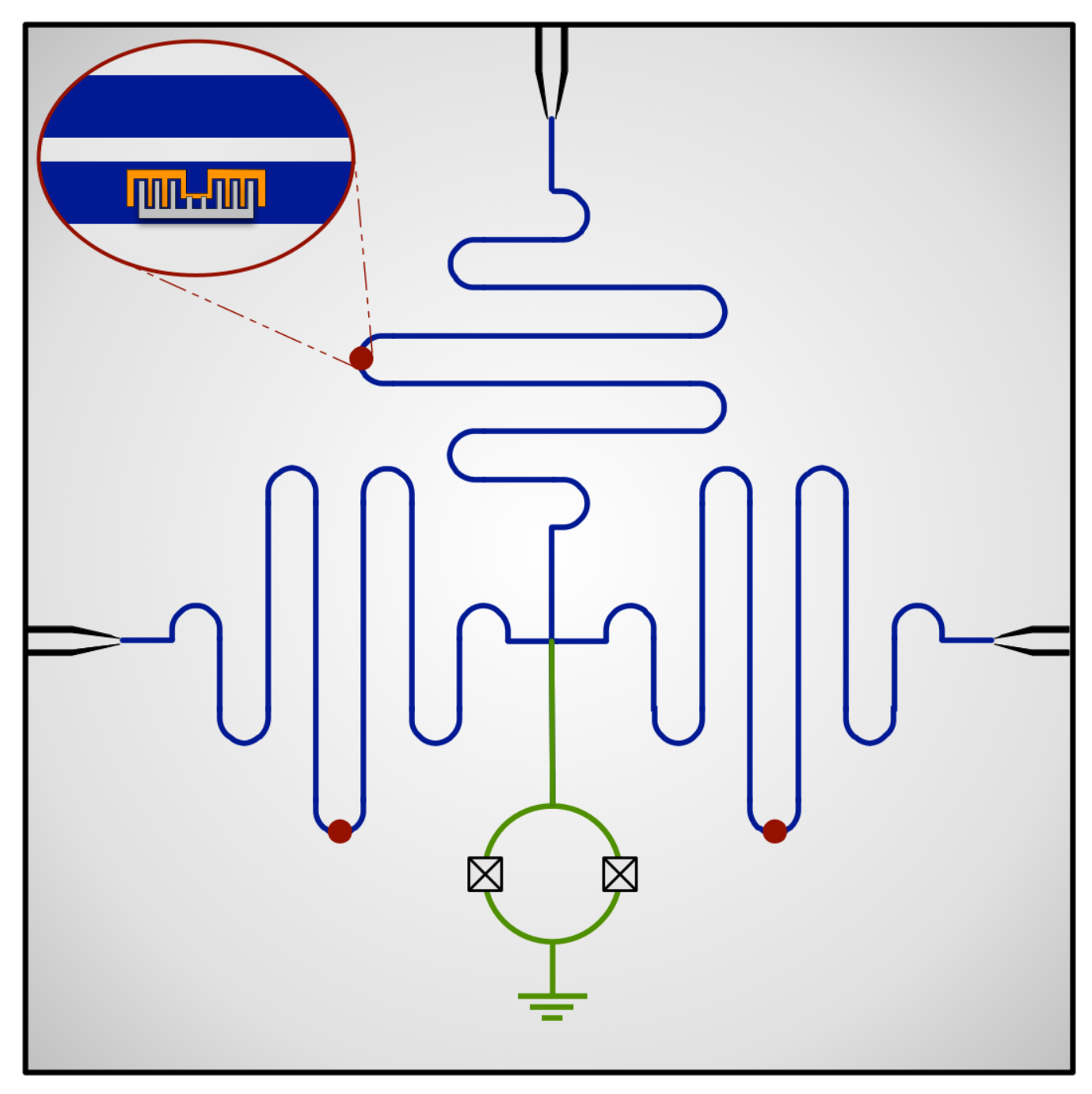}
\caption{\label{Fig4}  Three coplanar waveguide resonators are connected to the ground through a SQUID. Each resonator is coupled with a resonant transmon qubit. This scheme allows generation of GHZ-like entangled states, through a first-order process. By using this circuit design as a building-block, it is possible to explore more complex configurations and to build scalable cavity networks (see Supplemental Material). }
\end{figure}

In the framework of superconducting circuits, resonators can be linked together in unidimensional and bidimensional arrays to build networks of quantum cavities and superconducting devices. This enables us to envision more complex configurations which generalize the concept of dynamical Casimir effect to the multipartite case. Let us consider three resonators connected to the ground via a SQUID,
 as shown in Fig.~\ref{Fig4}. By injecting a fast-oscillating magnetic flux through the SQUID results in varying boundary conditions, which generate correlated photons pairs distributed in the three cavity modes. Such a configuration has no direct analogy with optical cavities, as opposed to the bipartite case. The Hamiltonian that describes the circuit  of Fig.~\ref{Fig4} is composed of three JC interactions and three time-dependent direct couplings between the field quadratures of each resonator pair
\begin{eqnarray}
\label{multihamiltonian}
\mathcal{H}& =&   \hbar\sum_{\ell =1}^{3}  \left[ \omega_{\ell} a_{\ell}^\dagger a_{\ell} + \frac{\omega^q_{\ell}}{2} \sigma_{\ell}^z + g_{\ell} \left( \sigma_{\ell} ^+  a_{\ell} + \sigma_{\ell}^- a_{\ell}^\dagger  \right) \right] \\
&+& \hbar \sum\limits_{\langle \ell,m \rangle}\alpha_{\ell m} (t) \left( a_\ell^\dagger + a_\ell\right) \left(a_m^\dagger + a_m \right). \nonumber
\end{eqnarray}
If the external flux threading the SQUID is composed of three signals oscillating at the frequencies $\omega_{\ell m}^d= \omega_\ell+\omega_m$, we can isolate the two-mode squeezing terms as in Eq.~\eqref{Twomode}.
 
\begin{figure}[t]
\centering
\includegraphics[angle=0, width=0.5\textwidth]{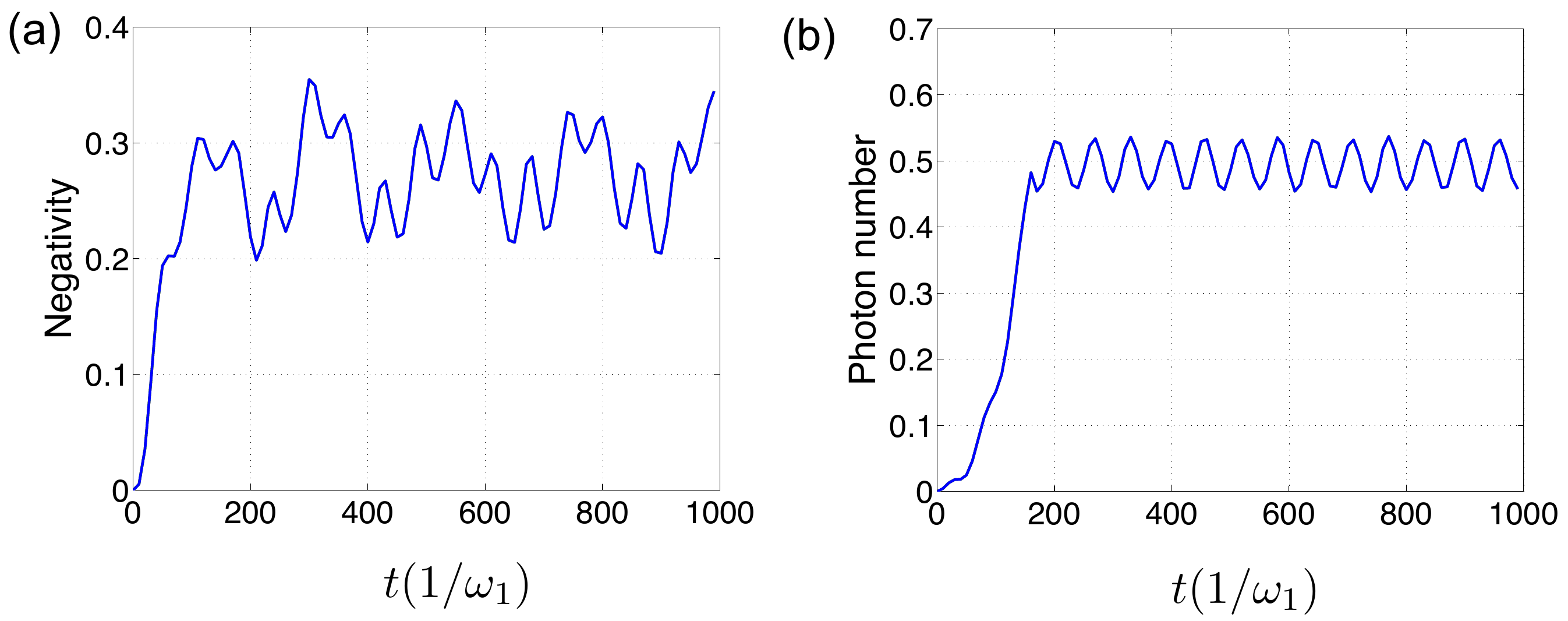}
\caption{\label{Fig5}  (a) Negativity  of the bipartite system obtained isolating one qubit from the set of the other two, as a function of time. Here, we considered resonator frequencies of $\omega_1/2\pi=3.8$~GHz, $\omega_2/2\pi=5.1$~GHz and $\omega_3/2\pi=7.5$~GHz. The SQUID is identical to the bipartite case and we use resonant qubits. The coupling parameters are homogeneous and their bare value is given by $\alpha_0 =5\ \omega_1\times 10^{-3}$.  (b) Average photon number in each cavity as a function of time. Due to the symmetric configuration the photon distribution is the same for the three cavities.}
\end{figure}

Generating multipartite entanglement is a challenging task, since it requires multiqubit gates whose operation fidelity is considerably lower than the single- or two-qubit gates.Here we show that our protocol allows generation of genuine multipartite entanglement (GME). With GME, we refer to quantum correlations which cannot be described using mixtures of bipartite entangled states alone.The negativity \cite{Vidal2002} is an entanglement monotone that estimates the bipartite entanglement shared between  two subsystems of any possible bipartition, it ranges from zero for separable to $1/2$ for maximally entangled states. It is defined  as $\mathcal{N}(\rho) = \frac{||\rho^{T_A}||_1 - 1}{2}$ where $||\rho^{T_A}||_1$ is the trace-norm of the partial transpose of the bipartite mixed state $\rho$. Numerical results on the negativity, shown in Fig.~\ref{Fig5}(a), indicate the generation of highly entangled states of three qubits. Figure~\ref{Fig5}(b) shows the average photon number in each cavity. In order to prove that such state is not biseparable, we evaluate an entanglement monotone that detects only multipartite quantum correlations, called  genuine multipartite entanglement (GME) concurrence $\mathcal{C}_{GME}$. It is obtained after an optimization process over all decomposable witnesses $W = P + Q^{T_A}$, where $P$ and $Q$ are positive semidefinite~\cite{Jungnitsch2011,Eltschka2012}.  Our results, $\max\left( \mathcal{C}_{\rm GME}\right)\approx0.3$, confirm the existence of genuine multipartite entanglement. 

Finally, to identify the entanglement class of three-qubit states, we make use of the entanglement witness~\cite{Kiesel2007} $\mathcal{W}_{\rm GHZ}  =3/4\ \mathbb{I} - P_{\rm GHZ}$, where $P_{\rm GHZ} = |\rm{GHZ}\rangle\langle\rm{GHZ}|$. Negative values for $ \text{Tr}\left[ \rho \mathcal{W}_{\rm GHZ}\right]$ imply that for any decomposition $\rho = \sum_j p_j \rho_j$ at least one $\rho_j$ is a GHZ state, and so $\rho$ belongs to the GHZ class. Local operations do not change the entanglement class, it means the witness can be optimized by minimizing $ \text{Tr}\left[ F \rho F^\dagger \mathcal{W}_{\rm GHZ}\right]$, where $F=F_1\otimes F_2\otimes F_3$, and $F_i$ are arbitrary single-qubit SLOCC operations. We obtained $\mathcal{W}_{\rm GHZ} = -0.06$, proving generation of (mixed) GHZ-like states, which belong to the most general entanglement class \cite{Acin2001}.

This scheme can be generalized to study entanglement generation in one- and two- dimensional cavity arrays in different geometries. Beyond the proposed model, our results show that superconducting circuits technology allows us to exploit the DCE physics as a useful resource for scalable quantum information protocols, generation of multipartite entanglement in artificial atoms, and as a building block for microwave quantum networks.

We thank G\'eza T\'oth and Jens Siewert for useful discussions. The authors acknowledge support from Spanish MINECO FIS2012-36673-C03-02; Ram\'on y Cajal Grant RYC-2012-11391; UPV/EHU UFI 11/55; Basque Government IT472-10;  CCQED, PROMISCE, SCALEQIT EU projects; and the Swedish Research Council.

\begin{widetext}

\section*{Supplemental Material}

In this supplemental material, we detail the derivation of the quantum model of the circuit design showed in Fig.~$1$b of the main text, and we briefly discuss possible future development of our work. In section \ref{sec1}, we derive the full quantum Hamiltonian that describes the bipartite configuration. In section \ref{sec2}, we show how to extend the model to the multipartite case, and how  our proposal can be used as a building block to implement highly correlated cavity networks for quantum information and quantum simulation.

\section{Quantum model}
\label{sec1}
In this section, we derive the quantum model of the circuit design proposed in Fig.~\ref{Fig2}(a) of the main text. We restrict to consider the bare resonators, an effective interaction with resonant qubits can be added at the end of the derivation. Let us consider a circuit composed of two transmission line resonators (TLS), connected to the ground through the same superconducting quantum interference device (SQUID), as shown in Fig.~\ref{Fig6}. A SQUID is a superconducting loop interrupted by two Josephson junctions (JJ). Here we take the two JJs that constitute the SQUID to be identical: under this assumption, the SQUID effectively behaves as a single JJ \cite{SQUIDBook}, namely, as a non-linear tunable inductance shunted by a small capacitance. We also assume that the JJs are such that their Josephson energy is much bigger than their  charge energy 
$E_J \gg E_C$. 
\begin{figure*}[t]
\centering
\includegraphics[width=\textwidth]{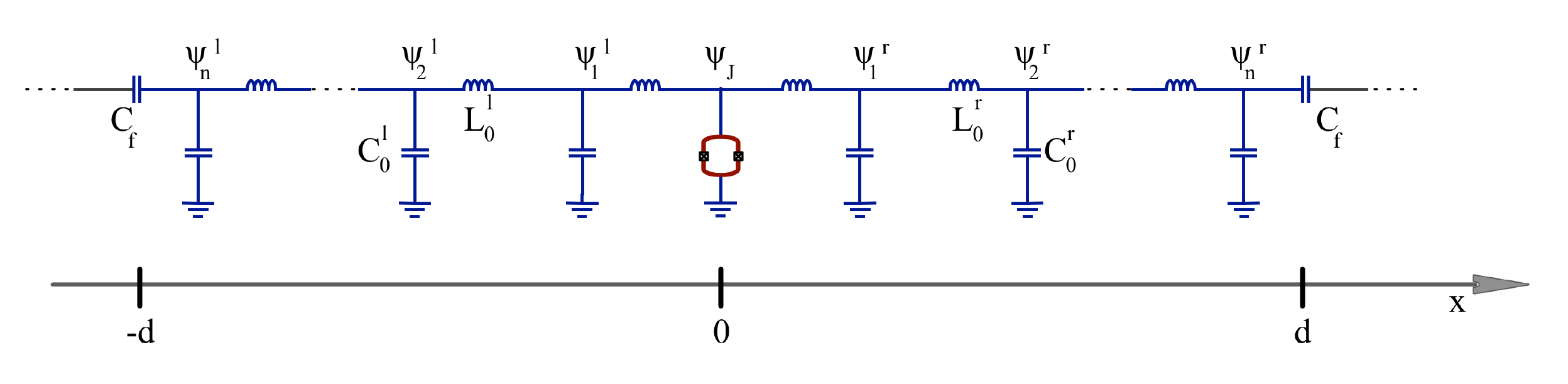}
\caption{\label{Fig6} \textbf{Sketch of the system.} Two transmission line resonators are connected to the same edge of a grounded SQUID. The SQUID low impedance imposes a voltage node at $x=0$. Each resonator is coupled with an external line (not considered here) needed for reading the cavity.}
\end{figure*}
In order to write the system classical Lagrangian, we will use a discrete description of the TLSs: each resonator will be represented by an infinite series of LC oscillators of infinitesimal length $\Delta x$. The  system Lagrangian can be then written as
\begin{eqnarray}
\mathcal{L} &=& \frac{1}{2} \sum_i \left\{ \Delta x C_0^l \psild ^2  + \frac{1}{\Delta x L_0^l} \left( \psi_{i+1}^l -\psi_{i}^l \right)^2  \right\} \label{freeleft} \\
&+& \frac{1}{2} \sum_i \left\{ \Delta x C_0^r \psird ^2  + \frac{1}{\Delta x L_0^r} \left( \psi_{i+1}^r -\psi_{i}^r \right)^2  \right\} \label{freeright} \\
&+& \frac{1}{2} C_J \para{\dot\psi_J}^2 - \frac{E_J\para{\phi_{\rm ext}}}{2\varphi_0^2}\psi_J^2 .
\label{lint}
\end{eqnarray}
We defined the magnetic flux $\psi_i^{l/r}$ in  the i-th inductor of the left/right resonator as the time integral of the instantaneous voltage $v_i$ across the element: $\psi^{l/r}_i (t)= \int\limits_0^t v_i (\tau)d\tau$.
The capacitance and inductance per unit of length are denoted by $C_0^{l/r}$ and $L_0^{l/r}$, respectively. Variables and constants with the subscript $J$ refer to the SQUID; notice that $C_J$ and $L_J$ represent the total capacitance and inductance of the SQUID, which will be described by means of a lumped-element model also in the continuum limit ($\Delta x \rightarrow 0$).
We defined the reduced magnetic flux as $\varphi_0 = \phi_0/2\pi$, where $\phi_0$ is the magnetic flux quantum. The inductance of the SQUID depends on the external flux $\phi_{\rm ext}$ threading the device: $L_J = \frac{\varphi_0^2}{E_J(\phi_{\rm ext})}$, where $E_J(\phi_{\rm ext}) = 2 E_J  \left|\cos{\para{\frac{\phi_{\rm ext}}{2\varphi_0}}}\right|$. The Josephson energy $E_J$ and the critical current $I_c$ are directly related 
$E_J = I_c \varphi_0$.

\subsection{Spatial modes}
\label{spatial}

In the bulk of each resonator the equation of motion is given by (for the sake of simplicity we omit the superscript \textit{l}/\textit{r})
\beq
C_0 \ddot{\psi}_i(t) = \frac{1}{\Delta x} \parc{\frac{\psi_{i+1}(t) -\psi_{i}(t)}{\Delta x L_0}-\frac{\psi_{i}(t) -\psi_{i-1}(t)}{\Delta x L_0}} 
\eeq
which, in the continuum limit $ \Delta x \rightarrow 0$, reduces to

\beq
\label{wave}
\ddot{\psi}(x, t) = v \left.\frac{\partial ^2}{\partial x ^2} \psi(x,t)\right|_{x=0} \qquad \text{where} \qquad v = \frac{1}{\sqrt{C_0L_0}} .
\eeq
The differential equation \eqref{wave} can be solved using the usual variable separation ansatz $\psi(x,t) = f(x) \phi(t)$, with $f(x)= \alpha \cos{\para{k x}} + \beta \sin{\para{k x}}$, 
$\phi(t) =  a e^{-i \omega t} + b e^{i \omega t}$, and $\omega = k/\sqrt{L_0C_0}$. 
The electrical boundary conditions at the far left  and far right extremities are established by the capacitances $C_f$, which mediate the coupling with external transmission lines. This capacitive coupling can be made very small and its contribution to the resonator modes is negligible.
Following a standard procedure, we will use open boundary conditions in order to evaluate the resonator modes, the interaction with the environment can be then described by means of a small effective coupling.
\beq
\label{open}
\left.\frac{\partial \psi^l(x)}{\partial x}\right|_{x=-d}=0 \qquad \text{and} \qquad \left.\frac{\partial \psi^r(x)}{\partial x}\right|_{x=d}=0 .
\eeq
The equation of motion for the dynamical variable $\psi_J$ corresponds to the Kirchhoff law of current conservation at the central node ($x=0$)
\beq
\label{kirchh}
C_J  \ddot\psi^2(0,t) + \frac{E_J\para{\phi_{\rm ext}}}{\varphi_0^2}\psi(0,t) = \frac{1}{L_0^l}\left.\frac{\partial \psi^l(x)}{\partial x}\right|_{x=o} + \frac{1}{L_0^r}\left.\frac{\partial \psi^r(x)}{\partial x}\right|_{x=0} .
\eeq
Now, given that the resonator inductances are much bigger than the SQUID inductance $L_J = \varphi_0^2/E_J\para{\phi_{\rm ext}}$, the terms on the right side of Eq.~\eqref{kirchh} are very small compared to the inductive contribution of the SQUID. 
 This mathematical statement has the following physical interpretation: the SQUID is a low-impedance element, therefore most of the current coming either from the left or from the right will flow  through the SQUID directly to the ground, without crossing
the other resonator. In other words, the presence of one resonator does not perceptibly modify the mode structure of the other one, although a small inductive coupling between them can be introduced from Eq. \eqref{kirchh}.
This allows us to define separated modes for the two resonators, which are constrained to satisfy the boundary condition
\beq
\label{bound}
\para{k\ d} \tan{\para{k\ d}} = \frac{C_J}{C} \para{k\ d}^2 - \frac{L}{L_J} .
\eeq
This equation holds for both resonators, we omitted the superscripts $l/r$ and we defined the total capacitance $C=C_0\ d$ and total inductance $L=L_0\ d$. Notice that modifying the external flux $\phi_{\rm ext}$ results in a variation of the SQUID effective inductance $L_J$, and so in a modification of the boundary condition \eqref{bound}.
Finally, the resonator frequencies can be found solving the differential equations \eqref{wave}, being the boundary conditions given by \eqref{open} and by the solution of the transcendental equation~\cite{Wall06} \eqref{bound}.

For didactical purpose, we observe that a good approximation to the resonator modes can be found assuming that at $x=0$ there is a voltage node, namely $\psi(0,t)=0$.
Under this assumption, the system is composed of two independent $\lambda/4$ resonators
\beq
f(x) = \sqrt{2}\sum_n \sin{\para{k_n\ x}} \qquad \text{where} \quad k_n = \frac{\pi}{d}\para{\frac{1}{2} + n} .
\eeq

\subsection{Hamiltonian}
Now, we write the system Lagrangian exploiting the stationary spatial solutions found in section \ref{spatial}. Integrating equations \eqref{freeleft} and \eqref{freeright} over the spatial degree of freedom, in the continuum limit, we obtain the Lagrangian of the free resonators
\beq
\label{l0}
\mathcal{L}_0 = \sum_{\nu=l,r}\sum_{n}\parc{\frac{C^\nu}{2} \parb{\dot\phi_n^\nu(t)}^2  - \frac{1}{L^\nu} \parb{\phi_n^\nu(t)}^2  } ,
\eeq
where the index $\nu$ identifies resonator, while $n$ runs over the spatial eigenmodes (and so, over the frequencies). We are interested in the dynamics of two level quantum systems that are embedded in the resonators. These qubits effectively interact only with one mode of each cavity, hence, hereafter we will restrict to consider one mode per resonator. This treatment is valid under the condition that the oscillation of the boundary conditions do not make resonant interaction terms between the relevant mode and the other ones.

The effective interaction between the modes $\phi^l$ and $\phi^r$ can be found isolating the variable $\phi(0,t)$ in Eq. \eqref{kirchh} and replacing it in the SQUID contribution to the system Lagrangian (Eq. \eqref{lint})
\beq
\label{lint2}
\mathcal{L}_{int} = -\frac{\varphi_0^2}{E_J\para{\phi_{\rm ext}}} \parc{ \frac{1}{L_0^l} k^l \phi^l(t)  + \frac{1}{L_0^r} k^r \phi^r(t)      }  ^2 .
\eeq
Now we assume that the prefactor of the previous equation is oscillating with a frequency such that only the cross-interaction term will be relevant in the Hamiltonian dynamics. This regime is accessible when the resonators are off resonance and the difference between their frequencies is much bigger than the coupling strength. We will show that the considered regime of parameters allows such approximation.
Defining the conjugate momentum $q^{l/r}=\partial \mathcal{L}_{tot}/\partial \dot\phi^{l/r}$, we find the system Hamiltonian
\beq
\mathcal{H} = \sum_{\nu=l,r}\parc{ \frac{1}{2C_\nu} q_\nu^2(t) +\frac{\omega_\nu^2 C_\nu}{2} \phi_\nu^2(t) } - \frac{2\varphi_0^2}{E_J\para{\phi_{\rm ext}}}\frac{\omega_l \omega_r}{Z_l Z_r} \phi_l \phi_r ,
\eeq
where we defined the impedance as $Z_\nu= \sqrt{L_\nu/C_\nu}$. Now, we perform the usual quantization process and define ladder operators
\beq
\parb{\phi_\nu, q_\nu}=i\hbar \qquad , \qquad
\phi_\nu = \sqrt{\frac{\hbar}{2\omega_\nu C_\nu}} \para{a_\nu^\dagger + a_\nu} \qquad , \qquad q_\nu = i \sqrt{\frac{\hbar C_\nu \omega _\nu}{2}} \para{a_\nu^\dagger - a_\nu} .
\eeq
Finally, we can write the  system  quantum Hamiltonian as
\beq
\label{theham}
\mathcal{H}/\hbar =  \omega_l a_l^\dagger a_l + \omega_r a_r^\dagger a_r - \frac{\varphi_0^2}{E_J\para{\phi_{\rm ext}}} \sqrt{\frac{\omega_l \omega_r}{C_l C_r}} \frac{1}{Z_l Z_r} 
\para{a_l^\dagger + a_l} \para{a_r^\dagger + a_r} .
\eeq

\subsection{Two-mode squeezing}
Observe that, as stated in the main text, when the driving frequency is comparable to the SQUID plasma frequency, the device can not be considered as a passive element and a more complex behaviour emerges.
The SQUID plasma frequency is defined as $\omega_p  = \sqrt{1/C_J L_J}$, so $\omega_p$ becomes smaller as the external flux get closer to $\phi_{\rm ext}/2\varphi_0=\pi/2$. 
To overcome this problem, we consider an external flux which is oscillating with small variation $\Delta$ around a fixed offset $\bar\phi$. In this way, with the physical parameters we considered, the SQUID plasma frequency is much bigger than $\omega_d$ for every value of $\phi_{\rm ext}(t)$ during the time evolution. Such condition allows us to expand the coupling parameter in the interaction
term of Eq. \eqref{theham}
\beq
\label{taylor1}
\frac{\phi_{\rm ext}}{2\varphi_0} = \bar\phi + \Delta \cos{\para{\omega_d t}} \qquad \Longrightarrow \qquad \frac{1}{E_J\para{\phi_{\rm ext}}} \approx \frac{1}{\cos{\bar\phi}} +
\frac{\sin \bar\phi}{\cos^2 \bar\phi}\ \Delta\ \cos{\para{\omega_d t}} .
\eeq
 Hence, the interaction term of the system Hamiltonian can be written, in the Schr\"odinger picture, as the sum of a constant and a time-dependent term
 \begin{eqnarray}
 \mathcal{H}/\hbar =  \omega_l a_l^\dagger a_l + \omega_r a_r^\dagger a_r + \eta\para{a_l^\dagger + a_l} \para{a_r^\dagger + a_r} + 
 \alpha_0 \para{e^{i\omega_d t} + e^{-i\omega_d t}} \para{a_l^\dagger + a_l} \para{a_r^\dagger + a_r} ,
 \end{eqnarray}
with
 \begin{eqnarray}
 \eta = \frac{\varphi_0^2}{2E_J \cos \bar\phi}\sqrt{\frac{\omega_l \omega_r}{C_l C_r}}\frac{1}{Z_l Z_r} \qquad \text{and} \qquad 
 \alpha_0 = \frac{\varphi_0^2}{4E_J}\frac{\sin \bar\phi}{\cos^2 \bar\phi}\sqrt{\frac{\omega_l \omega_r}{C_l C_r}}\frac{1}{Z_l Z_r} \Delta  .
 \end{eqnarray}
 When the detuning between the resonators is large compared to the coupling parameters $\eta, \alpha_0 \ll \left|\omega_l - \omega_r \right|$, we can perform the rotating wave approximation
 and neglect all terms that are  fast-oscillating in the interaction picture. If we choose the external driving to match the sum of the resonators frequencies $\omega_d = \omega_l + \omega_r$, the interaction Hamiltonian will reduce to a two-mode squeezing term
 \beq
\mathcal{H}/\hbar =  \omega_l a_l^\dagger a_l + \omega_r a_r^\dagger a_r + \alpha_0 \para{e^{- i\omega_d t} a_l^\dagger a_r^\dagger + e^{i\omega_d t} a_l a_r  } .
\eeq

\section{Multipartite case}
\label{sec2}
Consider a circuit scheme such that $n$ resonators are connected to the ground through the same SQUID, as shown in Fig. \ref{Fig8}(a) for $n=4$.
Equation \eqref{wave} still holds in the bulk of each resonator, and the boundary conditions \eqref{open} are still valid. On the other hand, the Kirchhoff's law of current conservation \eqref{kirchh} must be extended to include the contribution of every branch of the circuit
\beq
\label{multikirchh}
C_J  \ddot\psi^2(0,t) + \frac{E_J\para{\phi_{\rm ext}}}{\varphi_0^2}\psi(0,t) =\sum_{\nu} \frac{1}{L_0^\nu} \left.\frac{\partial \psi^\nu(x)}{\partial x}\right|_{x=o}  .
\eeq
As far as the resonator inductances are much larger than the SQUID inductance, we can still treat the system as composed of independent resonators, interacting through a small current-current coupling.
In this case, resonator spatial eigenmodes can be found following the same procedure we used in the bipartite case. Neglecting a small capacitive contribution, the term in the Lagrangian which describes the current-current coupling can be written as
\beq
\label{lint2}
\mathcal{L}_{int} = -\frac{\varphi_0^2}{E_J\para{\phi_{\rm ext}}} \parc{ \sum_\nu \frac{1}{L_0^\nu} k^\nu \phi^l(t)  }  ^2  .
\eeq
In a quantum description of such system, these interactions result in single-mode drivings and two-mode interactions between the field quadratures
\beq
\label{intterms}
\mathcal{H}_I = \sum_\nu \alpha_\nu(t) \para{a_\nu^\dagger + a_\nu}^2 + \sum_{\nu,\mu} \beta_{\nu \mu}(t) \para{a_\nu^\dagger + a_\nu} \para{a_\mu^\dagger + a_\mu},
\eeq
where the parameters $\alpha_\nu(t)$ and  $\beta_{\nu \mu}(t)$ depend on the external flux  $\phi_{\rm ext}(t)$ threading the SQUID, and they are all small compared to the resonator characteristic frequencies.
The time-dependence of $\phi_{\rm ext}(t)$ establishes which terms of equation \eqref{intterms} will have a non-negligible contribution to the system dynamics. When the external flux is given by the sum of signals oscillating at different frequencies, with small variations $\Delta_i$, around a constant off-set $\bar\phi$
\beq
\label{taylor1}
\frac{\phi_{\rm ext}}{2\varphi_0} = \bar\phi + \Delta_1 \cos{\para{\omega_{d1} t}} + \Delta_2 \cos{\para{\omega_{d2} t}} + \dots,
\eeq
with $\Delta_i\ll \bar\phi$, we can generalize the method used in equation \eqref{taylor1}
\beq
\frac{1}{E_J\para{\phi_{\rm ext}}} \approx \frac{1}{\cos{\bar\phi}} \ + \ 
\frac{\sin \bar\phi}{\cos^2 \bar\phi}\ \Delta_1\ \cos{\para{\omega_{d1} t}} \ + \ 
\frac{\sin \bar\phi}{\cos^2 \bar\phi}\ \Delta_2\ \cos{\para{\omega_{d2} t}}+ \dots
\eeq
Hence, controlling the external flux allows to turn on and off single- and two-mode squeezing terms, as well as linear couplings between the resonators.

\begin{figure*}[h]
\centering
\includegraphics[width=0.8\textwidth]{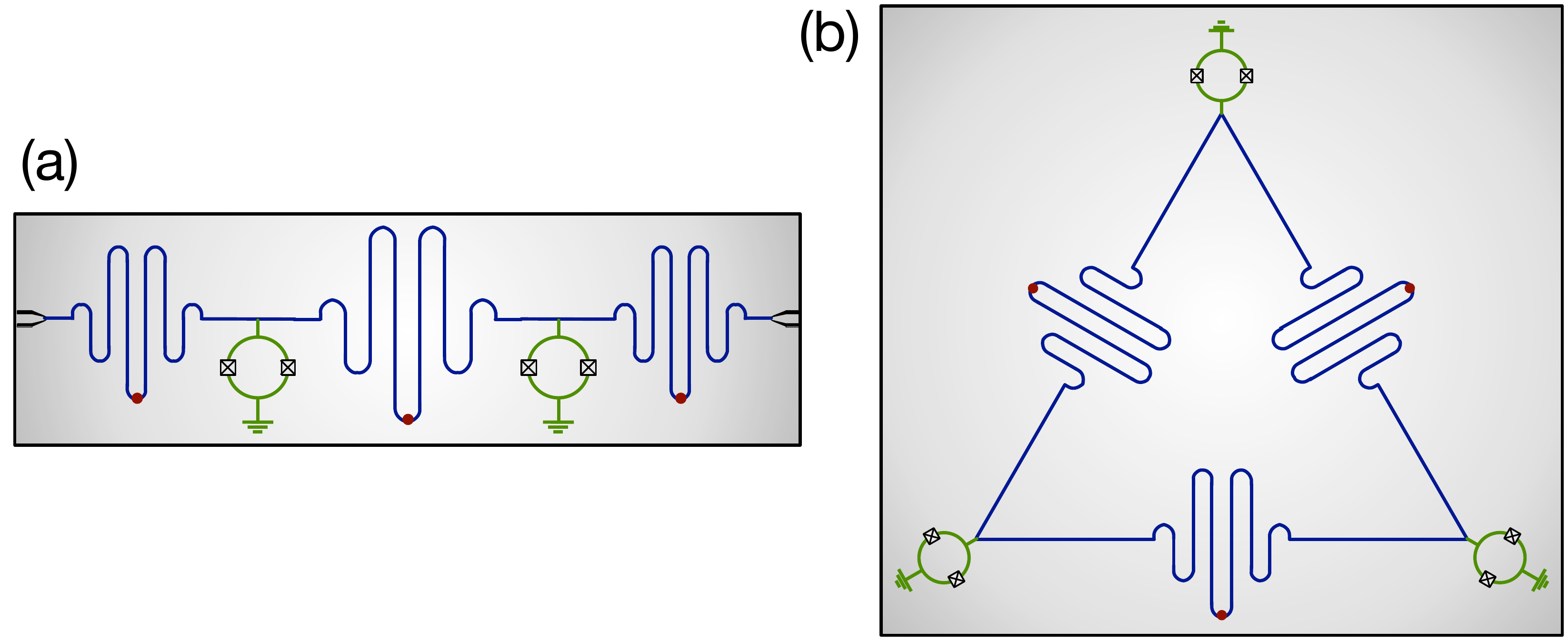}
\caption{\label{Fig7} \textbf{Tripartite setups.} {\bf (a)} Linear array of three resonators with near-neighbour couplings. {\bf (b)} Three SQUIDs in a triangular configuration. In this case, it is possible to control individually every current-current interaction in real time.}
\end{figure*}

\begin{figure*}[h]
\centering
\includegraphics[width=0.8\textwidth]{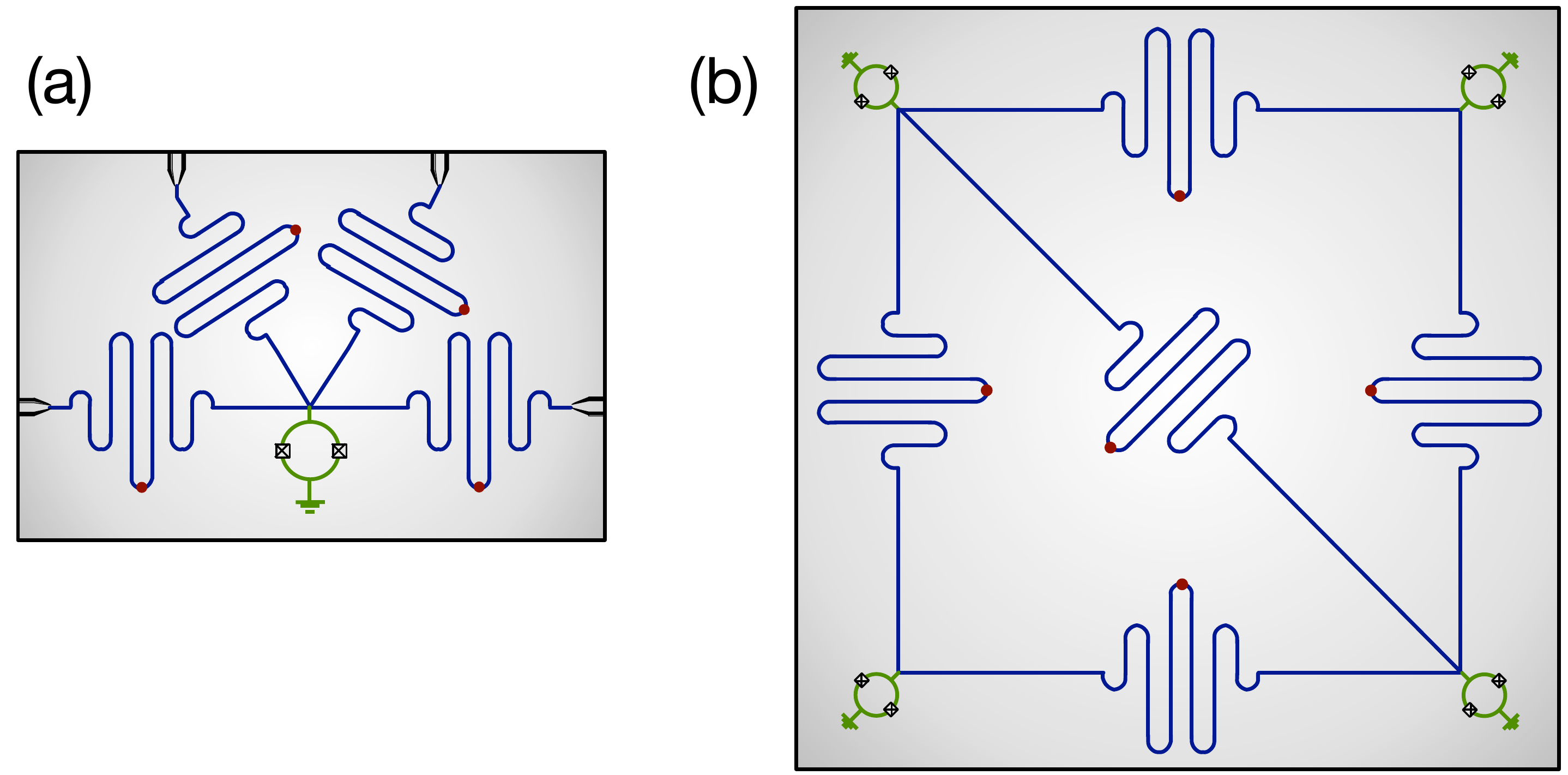}
\caption{\label{Fig8} \textbf{Multipartite case.}  Complex cavity configurations which deploy the dynamical Casimir physics in order to implement highly-correlated quantum networks.}
\end{figure*}

\subsection{Outlook}

Our theoretical analysis shows that fast-oscillating boundary conditions can generate a maximum entangled state of two qubits, in a non trivial way. This result demonstrates that the dynamical Casimir  effect (DCE) represents a valuable, so far overlooked resource for quantum information science. 
The implementation of quantum resonators ruled by fast-oscillating boundary conditions in superconducting circuits discloses the possibility of generalizing the dynamical Casimir physics to multipartite systems. Accordingly, we have theoretically proven that the DCE allows generation of three-qubit entangled states belonging to the GHZ class, i.e., to the most general class of genuine multipartite entanglement in the three-partite case.

Our proposal can be used as a building block to realize more complex circuit configurations, which exploit the dynamical Casimir physics in order to generate and distribute quantum correlations. Figure \ref{Fig7} shows two possible configurations of three-cavity setups: a linear array, box (a), and a triangular configuration, box (b). In the multipartite configuration presented in the main text (Fig. \ref{Fig4}), two-body interactions links resonators pairwise, while  the schemes of Fig. \ref{Fig7} lead to first-neighbour couplings. The linear array is interesting since it can be easily scaled to higher number of resonators. The triangular configuration allows real-time control of  the inhomogeneities in the couplings, due to the presence of three SQUIDs. Notice that, in the case in which both ends of a resonator are connected to a SQUID, the configuration of the spatial modes is such that there are voltage nodes at both extremities.
Figure \ref{Fig8}(a) shows a direct generalization of the three-partite scheme studied in the main paper, in which four cavities are involved. Such configuration is the most natural candidate to generate symmetric genuine multipartite entangled states of more artificial atoms. Finally, in Fig. \ref{Fig8}(b) it can be found an example of complex Casimir network, which shows the flexibility of the present proposal.

\end{widetext}
\end{document}